\documentstyle[12pt]{article}

\renewcommand{\thefootnote}{\fnsymbol{footnote}}
\begin{document}
\thispagestyle{empty}
{\hfill  Preprint JINR E2-93-194}\vspace{2.5cm} \\
\begin{center}
{\large{\bf Antibrackets and localization of  (path) integrals.}}
 \vspace{1.5cm}
 \\
{\large A. P.
Nersessian\footnote{E-MAIL:NERSESS@THEOR.JINRC.DUBNA.SU}}\vspace{0.5cm}\\
{\it Laboratory of Theoretical Physics,} \\
{\it Joint Institute for Nuclear Research}\\
{\it Dubna, Head Post Office, P.O.Box 79, 101 000 Moscow, Russia}
 \end{center}
\begin{abstract}
The transparent way for the invariant (Hamiltonian) description
of equivariant localization of the integrals over phase space is
proposed. It uses the odd symplectic structure, constructed over tangent
bundle of the phase space and permits straightforward generalization
for the path integrals. Simultaneously the method of supersymmetrization
for a wide class of the Hamiltonian systems is derived.
\end{abstract}

\begin{center}
{\it To appear in JETP Lett., {\bf 58} No.1 (1993)}
\end{center}

\vfill
\setcounter{page}0
\renewcommand{\thefootnote}{\arabic{footnote}}
\setcounter{footnote}0
\newpage
{\bf 1.} Recently a number of papers were published
( for example,~${}^{1- 3}$ ), where exact evaluation of the phase space path
integrals was studied  using corresponding generalization~${}^{1}$
of the Duistermaat---Heckman localization formula~${}^{4}$ (DH-formula).
In accordance with it,  if on the compact manifold $M$
provided with the symplectic structure
$\omega = \frac{1}{2\pi}\omega_{ij}dx^{i}\wedge dx^{j}$  the Hamiltonian
$H(x)$ defines the action of the group $U(1)\sim S^{1}$, then
\begin{equation}
Z_{0}=\int_{M}{\rm e}^{H} (\omega)^{N}=
\sum_{dH=0}\frac{{\rm e}^{H}\sqrt{det\omega_{ij}}}
{\sqrt{det\frac{\partial^{2} H}{\partial x^{i} \partial x^{j}}}} .
\label{eq:DH}  \end{equation}
Using its path integral generalization, one can localize the phase space
path integral into (finite-dimensional) integral over classical phase space.

This approach turns out convenient for a number of problems~${}^{2}$,
 in topological field theories particularly. It formed the basis for a
 conceptually new
 method of description of supersymmetric theories~${}^{3}$.

In the present letter we propose a simple method of the
invariant description of  DH-localization.
For this, following~${}^{1-3}$ we present the integral
(\ref{eq:DH}) in the form
\begin{equation}
Z_{0}=\frac{1}{(2\pi )^N}\int_{M}{\rm e}^{H(x)} \sqrt{det\omega_{ij}}d^{2N}x=
\frac{1}{\pi^N}\int_{{\cal M}}{\rm e}^{H-F}d^{2N}x d^{2N}\theta ,
\label{eq:int}  \end{equation}
where $\theta^i$ are auxiliary Grassmannian  fields ($p(\theta^{i})=
p(x^{i})+1$), which correspond to 1-forms $dx^i$, ${\cal M}$ is the
supermanifold associated with the tangent bundle
of $M$ ($z^{A}=(x^{i}, \theta^{i})$ are the local coordinates on $M$),
\begin{equation}
  F(z) =  -\frac{1}{2}\theta^i \omega_{ij} \theta^j .
\label{eq:F}\end{equation}
After that we shall define on ${\cal M}$  the {\it odd symplectic structure}.
The corresponding odd Poisson brackets (antibrackets )
give  the Hamiltonian description (and natural interpretation) of the
DH-localization without introduction of the additional structures, used in
the cited papers.

Besides we show that the use of antibrackets gives the
simple supersymmetrization method for the Hamiltonian systems,
which define the isometries of the Riemannian metric on the
their phase space.

Finally, the present constructions can be generalized straightforwardly
 to the case, if
$M$ is a symplectic {\it  supermanifold}. Moreover, they are
completely symmetrical according to the relation to initial
 and auxiliary coordinates.

All constructions presented in Letter relate to the
finite-dimensional integrals over compact symplectic manifolds.
One can accomplish their generalization for the path integrals
by the lifting on the loop space analogously~${}^{1-3}$.
It  does not  principally change the presented description scheme.

Notice that it is naturally connected with the Batalin---Vilkovisky
quantization formalism~${}^{5}$.\\

{\bf 2.}
Let us provide the supermanifold ${\cal M}$, which we defined above, with odd
symplectic structure
\begin{equation}
\Omega_{1} =\omega_{ij}dx^i\wedge d\theta^j
+\omega_{ij,k}\theta^j dx^i\wedge dx^k ,
\label{eq:osym}\end{equation}
where $\omega_{ij}$ corresponds to the symplectic structure on  $M$.

The corresponding to (\ref{eq:osym}) odd Poisson brackets ( antibrackets)
\begin{equation}
\{ f, g \}_1 = \frac{\partial_{r} f}{\partial z^A} \Omega^{AB}_{1}
\frac{\partial_{l} g}{\partial z^B}
\label{eq:bloc}
\end{equation}
are defined by the conditions:
\begin{equation}
\{x^i, x^j\}_{1}=0,\quad \{x^i, \theta^j\}_{1}= -\{\theta^{j}, x^{i}\}_{1} =
\omega^{ij},
\quad\{\theta^i, \theta^j\}_{1}=-\{\theta^j, \theta^i\}_{1}=
\frac{\partial\omega^{ij}}{\partial x^k}\theta^k
\label{eq:bxt}\end{equation}
where $\omega^{ij}\omega_{jk}=\delta^{i}_{k}$.
The antibrackets (\ref{eq:bloc}-\ref{eq:bxt} ) satisfy  the Jacobi identity:
   \begin{equation}
( -1)^{(p(f)+1)(p(h)+1)}\{ f,\{ g, h \}_{1}\}_1 +
{\rm {cycl. perm. (f, g, h)}} = 0 . \label{eq:bjac}
\end{equation}
Let us map the functions on  $M$ to the odd functions on ${\cal M}$ :
$$  f(x) \to Q_{f}(z) =\{f(x), F(z)\}_{1}, $$
where $F$ is defined by the expression (\ref{eq:F}).
 It puts  the Hamiltonian dynamics $\left ( H(x),\omega, M\right )$,
into odd one $\left ( Q, \Omega_{1}, {\cal M}\right )$,
where
\begin{equation}
Q=\{H, F\}_{1},
\label{eq:q}\end{equation}
with the equation of motion
\begin{equation}
\frac{dx^i}{dt} = \{x^i , Q\}_{1} = \{x^i , H_{0}\}_{0}\equiv\xi^i_{H} , \quad
\frac{d\theta^i }{dt} = \{\theta^{i} , Q\}_{1} =
\frac{\partial \xi^i_{H}}{\partial x^j}\theta^j .
\label{eq:motion}\end{equation}
This dynamics is supersymmetric: from the closeness of $\omega$ follows
 $\{F, F\}_{1}=0$,
and taking into account (\ref{eq:q})  we obtain the simplest superalgebra
\begin{eqnarray}
 && \{H\pm F, H \pm F\}_{1} =\pm 2Q , \label{eq:sualg1}\\
 && \{H+ F, H -  F \}_{1} = \{H\pm F, Q \}_{1}= \{Q, Q \}_{1} = 0 . \nonumber
 \end{eqnarray}
The following correspondence is obvious:
\begin{eqnarray}
&\{H,\quad\}_{1}&=\xi_{H}^i \frac{\partial}{\partial \theta^i}
\rightarrow \imath _{H} -{\rm the \quad operator \quad of \quad interior
\quad product
\quad on}\quad \xi_{H} ;\nonumber\\
& \{F,\quad\}_{1}&= \theta^i \frac{\partial}{\partial x^i}   \rightarrow
d -{\rm  the\quad operator\quad of\quad exterior\quad
differentiation};\nonumber\\
& \{ Q,\quad \}_{1}&=\xi_{H}^i \frac{\partial}{\partial x^i} +
\xi_{H,k}^{i}\theta^k \frac{\partial}{\partial \theta^i}
\rightarrow {\cal L}_{H} -{\rm the\quad Lie\quad derivative\quad along}
\quad \xi_{H} .\nonumber\\
\label{eq:corr}\end{eqnarray}
Taking into account the Jacobi identity (\ref{eq:bjac}) we have:
$$\{H, F \}_{1} =Q \rightarrow
d\imath_{H} +\imath_{H}d = {\cal L}_{H} -
{\rm homotopy\quad formula }.\nonumber$$
As we see, the supersymmetry of  $\left ( Q, \Omega_{1}, {\cal M}\right )$
corresponds to the equivariant differentiation $d_{H}=d+\imath_{H}$.

Following the papers~${}^{1,2}$, let us assume that on $M$
the Riemannian metrics $g_{ij}$ which is Lie-derived
with $\xi^{i}_{H}$ is defined also.
Then the odd function
\begin{equation}
{\tilde Q} = \xi^i_{H} g_{ij}\theta^j \equiv \xi_{i}\theta^i
\label{eq:gauge} \end{equation}
is the  integral of motion of (\ref{eq:motion}):
\begin{equation}
{\cal L}_{H} g = 0 \rightarrow\{Q, {\tilde Q}\}_{1}=0 .
\label{eq:killing}\end{equation}
We have also:
 $$  \{F, {\tilde Q} \}_{1}= - F_{2} ,\quad \{H, {\tilde Q} \}_{1} = H_{2}
$$
where
$$ H_{2}= \xi^{i}_{H}g_{ij}\xi^{j}_{H},\quad
F_2 =\frac{1}{2}\theta^{i}\omega_{(2)ij}\theta^j,
\quad \omega_{(2)ij}=\frac{\partial \xi_{i}}{\partial x^j} -
\frac{\partial \xi_{j}}{\partial x^i}. $$
\\
{\bf 3.} Now we shall demonstrate the derivation of DH-formula
 (\ref{eq:DH}) using the constructions presented above.

Let us consider the integral
\begin{equation}
Z_{\lambda}=\frac{1}{(\pi)^N}\int_{{\cal M}}\exp(H-F-
\lambda \{H+F, {\tilde Q\}_{1}}) d^{4N}z ,
\label{eq:int2} \end{equation}
where $\lambda$ is arbitrary numerical parameter.

As in the first item, we assume that  ${\cal M}$ is associated with the tangent
bundle
of the compact symplectic manifold $M$, and  define on it the odd
symplectic structure (\ref{eq:osym}). We also assume that the Hamiltonian
$H(x)$ defines on $M$ the action of $U(1)\sim S^{1}$, that  $M$ provided
with the Riemannian structure $g_{ij}$, which is Lie-derived
with  $\xi^{i}_{H}$,
and that $F$ and ${\tilde Q}$ define by the expressions (\ref{eq:F}) ,
(\ref{eq:gauge}).

The vector fields (\ref{eq:corr}) preserve the ``volume form"  $d^{4N}z=
d^{2N}x d^{2N}\theta $.
 From (\ref{eq:sualg1}), (\ref{eq:killing})
we deduce
$$\{H+F , {\rm e}^{H-F-\lambda \{H+F, {\tilde Q}\}_{1}} \}_{1}=0,\quad
\{Q , {\rm e}^{H-F-\lambda \{H+F, {\tilde Q}\}_{1}} \}_{1}=0. \nonumber  $$
Therefore the integral (\ref{eq:int2}) is invariant under
equivariant and Lie transformations along $\xi^{i}_{H}$.
We have also
$$\{Q , {\tilde Q}{\rm e}^{H-F-\lambda \{H+F, {\tilde
Q}\}_{1}}\}_{1}=0.\nonumber$$
Using  these expressions and the fact that the integral of an equivariantly
exact form vanishes, we show that
\begin{eqnarray}
\frac{dZ_{\lambda}}{d\lambda}
&=&-\lambda\frac{1}{\pi^N}\int_{\cal M}\{H+F, {\tilde Q}\}_{1}{\rm e}^{H -F-
\lambda \{H+F, {\tilde Q}\}_{1}} d^{4N}z =  \nonumber \\
&=&-\lambda\frac{1}{\pi^N}\int_{\cal M}\{H+F, {\tilde Q}{\rm e}^{H -F-
\lambda \{H+F, {\tilde Q}\}_{1}}\}_{1} d^{4N}z +\nonumber\\
&+&\lambda\frac{1}{\pi^N}\int_{\cal M}{\tilde Q}\{H+F, {\rm e}^{H -F-
\lambda \{H+F, {\tilde Q}\}_{1}}\}_{1} d^{4N}z =0 \nonumber
\label{eq:lambda}\end{eqnarray}
 Thus, provided the limits $\alpha\to 0$, $\alpha\to\infty $ and
taking into account that
$$\delta (\xi^{i}_{H})= \frac{1}{\pi^{2N}}\lim_{\lambda\to\infty}
\sqrt{\lambda^{2N}det g_{ij}}
{\rm e}^{-\lambda \xi^{i}_{H}g_{ij}\xi^{j}_{H}},\nonumber$$
 we obtain the DH-localization formula:
\begin{eqnarray}
&Z_{0}&=\frac{1}{(2\pi)^N}\int_{M}{\rm e}^{H}\sqrt{det\omega_{ij}}d^{2N}x=
\lim_{\alpha\to\infty}\frac{1}{\pi^N}\int_{{\cal M}}{\rm e}^{H -F-
\lambda (H_{2}-F_{2})}d^{4N}z = \nonumber\\
&&=\int_{M}{\rm e}^{H}\delta{(\xi_{H})}\sqrt{\det\omega_{ij}}\sqrt{\det
\frac{\partial \xi_{H}^{i}}{\partial x^j}}d^{2N}x. \nonumber
\label{eq:DH2} \end{eqnarray}
Generalization of the presented constructions
 for the path integrals  can be accomplished
by the lifting on the loop space similarly to~${}^{1-3}$.

Then $H\rightarrow \int{A_{i}dx^{i} - H dt}$ ( where $dA=\omega$) ,
$\xi^{i}_{H}\rightarrow \xi^{i}_{S}=(\dot{x}^{i} - \xi^{i}_{H}) $, and
the path integral localizes in the ordinary integral over the
classical phase space.

Note that the representation of the initial integral in the
form (\ref{eq:int})  formally coincides with the form of the integral from
differential forms in the case where $M$ is the {\it supermanifold}~${}^{6}$.
Note also, that the present description is symmetric according to
 the relation to initial
 and auxiliary  coordinates.
Then it can be generalize for the super-Hamiltonian systems.\\

{\bf 4}.
If on $M$ both symplectic and Riemannian structures
are define, then on ${\cal M}$ one can  also construct {\it even}
{\it symplectic } {\it structures}
$${\Omega}_{\alpha}=
\frac{1}{2} (\omega_{(\alpha)ij} +
R_{ijkl}\theta^{k}\theta^{l})dx^{i}\wedge dx^{j} +
g_{ij}D\theta^{i}\wedge D\theta^{j}, \quad \alpha=0,2  $$
where
$D\theta^{i} = d\theta^{i} + \Gamma^{i}_{kl}\theta^{k}dx^{l}$ ;
$R_{ijkl} $ ,  $\Gamma^{i}_{kl}$ -- correspondingly curvature
and connection associated with the metric $g_{ij}$ on $M$,
$\omega_{(0)ij}\equiv\omega_{ij}$.

It is easy to see  that    $( H_{0}+F_{2}, {\Omega}_{0},
{\cal M})$, $( H_{2}+F_{2}, {\Omega}_{2},
{\cal M})$ and $(Q, \Omega^{1} , {\cal M})$,  define the same supersymmetric
dynamics (\ref{eq:motion}), if $g_{ij}$ is Lie-derived with $\xi^{i}_{H}$.

The example of supersymmetric dynamics provided
with both even and odd Hamiltonian structure
(one dimensional Witten dynamics) was  first proposed
by  D. V. Volkov et al.~${}^{7}$ . Such dynamics were considered
also in~${}^{8}$.\\
\newpage
{\large {\bf References}}\\
${}^{1}$M. Blau, E. Keski- Vakkuri, A.J. Niemi
- Phys.Lett., {\bf 246B}, 92  (1990)\\
${}^{2}$ A. J. Niemi, P. Pasanen  - Phys.Lett., {\bf 253B}, 349 (1991)

 A. J. Niemi, O. Tirkkonen  - Phys.Lett., {\bf 293B}, 339 (1992);

A. Hietaki, A. Yu. Morozov, A. J. Niemi, Palo K.- Phys. Lett. {\bf B263}, 417
(1991) \\
${}^{3}$  A. Yu. Morozov, A. J. Niemi, K. Palo - Phys. Lett. {\bf B271}, 365
(1991);
 Nucl. Phys.

 {\bf B377}, 295 (1992)\\
${}^{4}$ J. J. Duistermaat, G. J. Heckman - Inv. Math. {\bf 69}, 259 (1982);
 {\it ibid} {\bf 72}, 153 (1983) \\
${}^{5}$ I. A. Batalin, G. A. Vilkovisky - Phys.Lett., {\bf 102B}, 27 (1981);
Nucl.Phys., {\bf  B234}, 106

(1984) \\
${}^{6}$ I. N. Bernstein, D. A. Leites - Funct. Anal. Appl. {\bf 11}, No. 2, 70
(1977)\\
${}^{7}$ D. V. Volkov, A. I. Pashnev, V. A. Soroka, V. I. Tkach -
 JETP Lett. {\bf 44},

55 (1986)   \\
${}^ {8}$ O. M. Khudaverdian, A. P. Nersessian  - J. Math. Phys., {\bf 32},
1938 (1991) ;
Preprint

JINR E2-92-411;

A. P. Nersessian - Preprint JINR P2 - 92 -265 (in Russian ),
Theor. Math. Phys.,{\bf 96} No. 1

(to appear )

\end{document}